# Implicit Location Sharing Detection in Social Media from Short Turkish Text

Davut Deniz YAVUZ
davut.deniz.yavuz@gmail.com
Department of Computer Engineering,
TOBB University, Ankara

Osman ABUL
osmanabul@etu.edu.tr
Department of Computer Engineering,
TOBB University, Ankara

**ABSTRACT**

*Social media have become a significant venue for information sharing of live updates. Users of social media are producing and sharing large amount of personal data as a part of the live updates. A significant percentage of this data contains location information that can be used by other people for many purposes. Some of the social media users deliberately share their own location information with other social network users. However, a large number of social media users blindly or implicitly share their location without noticing it or its possible consequences. Implicit location sharing is investigated in the current paper.*

*We perform a large scale study on implicit location sharing for one of the most popular social media platform, namely Twitter. After a careful study, we built a dataset of Turkish tweets and manually tagged them. Using machine learning techniques we built classifiers that are able to classify whether a given tweet contains implicit location sharing or not. The classifiers are shown to be very accurate and efficient. Moreover, the best classifier is employed as a browser add-on tool which warns the user whenever an implicit location sharing is predicted from to be released tweet. The paper provides the methodology and the technical analysis as well. Furthermore, it discusses how these techniques can be extended to different social network services and also to different languages.*

## 1. INTRODUCTION

The majority of social media users are not aware of the risks when sharing personal information in social media. A lot of people share large amount of data for having good time and being recognized in social media. People share photos, text messages, location, and expressions in social media venues such as Twitter, Facebook, and Myspace; whereas another group of people are eager to use this essential information for advertising, marketing, and many other purposes. Managers use this information to chase their employee; they make promotion decisions within employers by also considering their social media updates. Managers also use this information to evaluate candidates when hiring. It has been stated that an October 2011 Society for Human Resource Management (SHRM) survey of more than 500 of its members involved in recruiting found that about 18 percent use social network searches to screen candidates [1]. Even worse, some people use this information for choosing their victims. For instance, thieves use location information to determine if a house is empty and available for robbing. pleaserobme.com and geosocialfootprint.com are examples of how easy to find people's exact location.

There are literature studies and applications developed on providing location privacy; some of which are based on free text documents. However, they are generally available for English only and their semantic analysis is in English too. They mostly aim to raise awareness on how unsafe to share location in social media, they do not propose a research to prevent or notify users before sharing their location.

In this study, we focus on Turkish tweets and extracted location features based on Turkish language. Furthermore, we developed a Google Chrome extension that notify users before they send a tweet on Twitter.

### 1.1 Location Sharing on Social Media

Since many people are following and looking each other's social pages constantly. For instance, social media is becoming the first place that comes to people' mind to locate where a person is. In Facebook, Twitter and Foursquare people share great amount of data, sizeable portion of which is personal data including status, text messages, photos, videos and many other kinds of updates.

With GPS enabled mobile devices people can share location with fine granularity and this potentially leads to location privacy leaks. Intruders may use personal location information in many ways, the most innocent of which is advertising. Consider the scenario, suppose you check-in in a mall by your social media profile and some companies use your location information and send you a targeted advertisement via the form of e-mail or SMS, offering discount in a certain store. This can be considered a location privacy violation with low risk. On the other hand, the main risk, however, the disclosed location information can be used to track the victim and make physical assaults. In Foursquare, one can add the his home address as a place or even worse other peoples can add the home as the current location, and people can make check-ins in the home. This may lead to friends of friends to know the home address. It is possible to find people's home location by 77.27% accuracy in a radius less than 20km from Foursquare [2]. This can give rise to many location related risks.

There are some precautions to prevent users from these kind of risks. In Foursquare, users can send an e-mail to privacy@foursquare.com about deleting their home address from database [3]. Also, users can adjust their privacy settings so that only close friends can see check-ins. In Facebook and Twitter there are also privacy settings to protect users' location privacy.

However, most of the people neglect these settings and are not aware of the risks of location sharing. This is because maintaining privacy settings in social media is not always what people really like to do. They rather tend to use default settings as they think they would damage their account settings. Even the people make appropriate location privacy settings, the location sharing risk are still on the table. Clearly they may type their location directly or indirectly and implicitly provide information that can be used to link to users' whereabouts. Hence, social media users need a preventive way that interactively check their posts and determine if the post shares a private location information.

## 1.2 Location Sharing on Twitter

Users of Twitter can share their location with many different ways. For mobile phones, people can add a location label and use GPS to give their exact locations. When tweeting location on a device, one can also easily enable location sharing. Another way to share location is to use other location applications such as Foursquare. People can link their Twitter accounts to Foursquare in order to automatically tweet their location.

In this study, we examined that 20.903 of 537.125 tweets use Foursquare and 5.567 tweets use Twitter's feature to share location in Turkey. In the pie chart we can see that approximately 5% of tweets use explicit location sharing and 79% of them use Foursquare to share their location. People that use these technologies are aware of sharing their location and they share location on purpose.

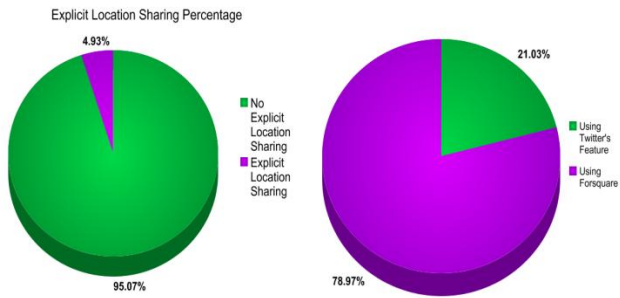

**Figure 1: Pie chart of explicit location sharing over all tweets collected and the percentage of the technologies used for explicit location sharing**

However, a group of people on twitter are not aware when sharing their location. This happens when they do not use Foursquare or Twitter's location sharing option. Some people send tweets blindly and they do not know they share location. For example, a tweet that says; 'Çok sıkıldım, evde yapıcak bir şey yok.' which means 'I am very bored, nothing to do at home.' share location information that the user is at home at that time, but the user may not notice sharing location. Another example to that is 'Armada'da gezecek çok mağaza var, gezerken çok yoruldum.' which means 'There are a lot of shop in Armada, I got tired while walking around.' in English. In this tweet the user implicitly shares he is in the famous shopping mall Armada in Ankara. In Map 1 you can see how easy to find that the person's exact location by using Google Maps. These examples are hard to notice for location sharing for users because the main purposes of these tweets is to talk about how tired the person is or how boring is the time, not sharing location to other users. These types of tweets are within the scope of implicit location sharing where it is the main concern of our study.

**Map 1: The address of Armada from Google Maps**

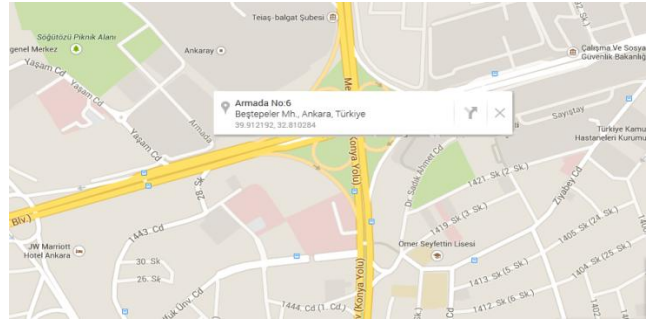

## 2. Related Work

Over the last several years, many researchers have developed ideas and done studies about social network privacy and location privacy on social media. Several researches aim to raise awareness about location privacy or to find peoples location by collecting the data from social network. For example, GeoSocial Footprint is an online tool that provides tweeter users with an opportunity to view their geosocial footprint [4]. This tool gets any twitter user name and gives you a map of where the user was in the past. It also gives you some suggestions on how to decrease your geosocial footprint.

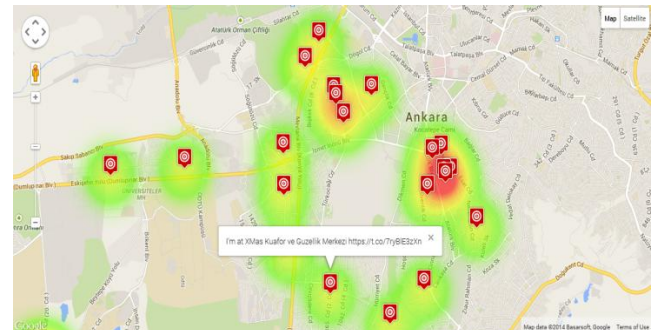

**Figure 2: Example to the result map of geosocialfootprint.com**

Figure 2 is an example of how one can view the user's location history using this tool. You can also click to the location icon and see the tweet that share the location. Using the tweet you can also determine the time of the tweet by looking at the users timeline. It is an undeniable fact that this study will increase the awareness on location privacy, on the other hand, some malicious people may use this location information to harm you.

This site is based on a research which is published by *Weidemann* in the paper "Social Media Location Intelligence: The Next Privacy Battle - An ArcGIS add-in and Analysis of Geospatial Data Collected from Twitter.com [5]". In Weidemann's study, it is stated that GISsience professionals are aware of the potential risks when using social network, but the general public usually do not know about these risk, it is also said that the real time record of the people's location is more treasured information than credit card numbers and bank statements. The findings of this study shows that 0.8 percent of Twitter users share their current location via GPS or other technologies and 2.2 percent of all tweets provide ambient location data. We can conclude from these findings that twitter users share enough data to people who want use their location information for whatever purpose they desire [5].

Another example of location awareness is http://pleaserobme.com. This web site gives others the opportunity to check your own timeline for check-ins. The creators of this web site, Frank Groeneveld, Barry Borsboom and Boy van Amstel, mention that, social networks have great searching engines for users. By these functionalities people find their friend and the things they are interested. However, if you allow your messages to travel between different social networks it becames more complicated to track your privacy and information you trust to your friends might end up somewhere else [6]. For instance if you link your foursquare account to your twitter account your privacy settings in foursquare will not work on your twitter account and you can not protect your foursquare location information from other public Twitter users. (Twitter: "Our default is almost always to make the information you provide public" [7])

It is clear that, those tools and researches increase the awareness of people about location sharing. However, very limited researches have done to prevent social media users form implicit location sharing. In this study, we examined the Turkish language grammar properties and used data mining algorithms to prevent twitter users from sharing location before they send their tweets.

Sentiment analysis is also popular from tweets and recently the research by Taşçıoğlu shows it is possible to detect irony in Turkish micro block texts [8]. It uses sentimental analysis techniques, data mining algorithms and natural language processing to determine the irony within statements. In the study, Taşçıoğlu used Twitter API to collect various Turkish tweets and compare different classification algorithms to determine the most suitable classifier. We use similar machine learning techniques to detect location sharing in Twitter.

## 3. Methodology

Stefanidis presents three components of system architecture for collecting information from social media feeds, which are; extracting data from the data providers via APIs; parsing, integrating, and storing these data in a resident database; and then analyzing these data to extract information of interest [9]. In this study we also follow a similar system to achieve what we intended. The first step is the data collection phase. In this step we collected tweets via Twitter4j. The next step is constructing the data set. The data set is created manually by looking at the collected tweets. After this step we came to the feature extraction phase. In this phase features are extracted by considering Turkish grammar properties. To find classifiers in the form of decision tree for the selected features we used Weka. After selecting the classifier and fixing the best decision tree we developed a Google Chrome Extension to check tweets by considering the decision tree. The flow chart of these steps is shown in Flow Chart 1.

## 4. Data Set and Feature Extraction

### 4.1 Data Collection

Twitter4j [10] which is a Java library for the Twitter API [11] is used to collect tweets from various tweeter users who lives in Turkey. 537127 tweets are collected from 1813 distinct users. These tweets are in Turkish and there is no certain classification information in order to have different type of tweets in different subjects. To store the collected tweets and to work on these tweets MySQL [12] database management system is used.

### 4.2 Data Set

500 tweets subset from 537127 tweets are carefully selected to form the data set. 250 ouf of 500 of them implicitly sharing location information are manually marked in the database.

Similarly, the remaining 250 tweets which do not implicitly share location are also manually marked in the database too. Since the purpose of this study is to semantically analyze the tweets for location privacy, we exclude tweets which use foursquare and similar technologies to share location information.

Tweets that contain a city name, a specific place like restaurant, home and university are marked for sharing location. On the other hand, tweets that talk about politics, football are considered as they do not share location and they are marked as no location sharing.

### 4.3 Feature Extraction

By considering the data set and the Turkish language, 6 features are extracted. Although location sharing attribute is marked manually, features are marked by SQL statements automatically when existence of the feature is checked. When deciding on the features, TDK dictionary [13] is used for finding location indicative words and verbs. Index-Anatolicus [14] is also used for listing the city names in Turkey.

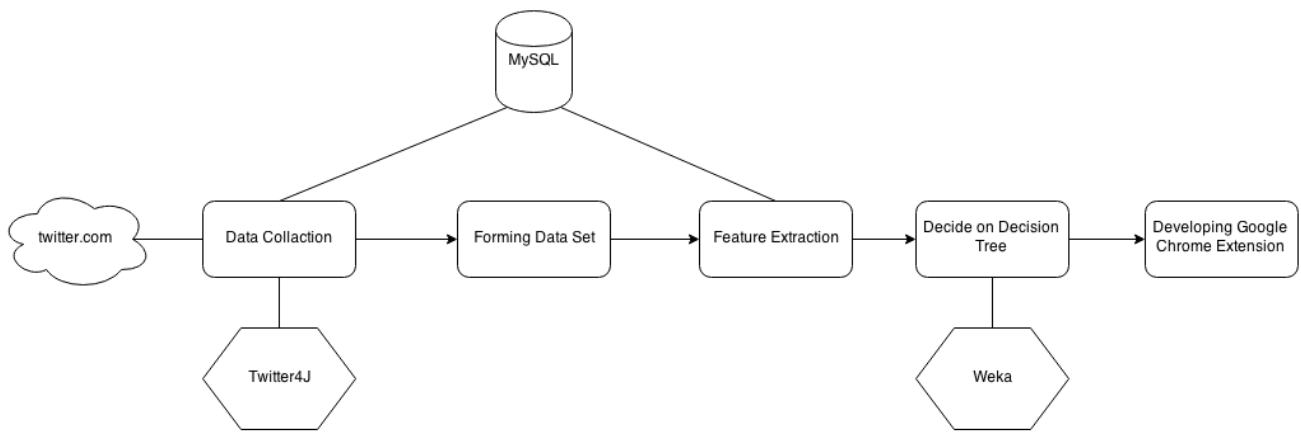

**Flow Chart 1: Methodology**

### 4.3.1 Feature 1
Feature 1 consists of two Turkish suffixes which are 'deyim' and 'dayım'. These means 'I am at' in English. For example, a tweet that says; 'Sınavlar yüzünden bugün yine okuldayım' which is in English, 'Because of the exams I am at school once again' shares location information. Another example to that; 'Hastalığım yüzünden evdeyim, dışarı çıkamıyorum' which means in English 'I am at home and I can't go outside because of my illness.' Also shares location information.

4 tweets found in the data set which has this feature and two of them share user's location. The other two are found to be non-location sharing suffixes.

### 4.3.2 Feature 2
Feature 2 is formed by two Turkish suffixes that are 'de' and 'da'. These two suffixes are searched at the end of the words. These words means like 'at' and 'in' in English. To exemplify, a tweet like this; 'Okulda sınıf çok sıcak' which means 'Class is too hot at the school' shares location information. A second example to that; 'Terminalde arkadaşımı bekliyorum.' In English 'I am waiting my friend at the terminal' also shares location.

158 tweets are marked for this feature. 97 of them are marked as location sharing statements.

### 4.3.3 Feature 3
Some special words are chosen to form feature 3 such as okul, ev, iş, cafe, kafe, restoran. 48 words are used in this feature. These words are significant words that have a great potential of location indicators when they are used in a sentence. One example is 'Şu anda Avlu restorana gidiyorum' which means 'I am going to Avlu restaurant right now' is sharing where you will be soon. Another example is 'Ev çok dağınık' which means 'The house is so messy' is sharing where you are.

248 tweets are found to ensure the feature 3. 191 of these tweets have location sharing.

### 4.3.4 Feature 4
City names in Turkey are used for feature 4. There are 81 cities in Turkey, and all of them are included in feature 4 without case sensitivity. For example, 'Ankarada yapacak hiçbir şey yok.' Which means 'Nothing to do in Ankara' in English and 'Izmirde denize girmeyi çok özlemişim' which means 'I miss go swimming in Izmir very much' shares location information. There are 178 tweets that have feature 4 and 158 of them share location.

### 4.3.5 Feature 5
As discussed above in section 3.2, tweets that are built up with foursquare and other technologies similar to foursquare are not included in the data set, but they are used to create feature 5. Tweets that are created by these programs are similar to: 'I'm at Marco Pascha in Ankara, Türkiye https://t.co/HoJEeqXBhK'. We get the part 'Marco Pascha' and use it as a special word in feature 5. Because all the special words in these tweets are place names of cafes, restaurants, schools, universities, stadiums etc. We extract 6560 distinct place names and used for feature 5. For instance, a tweet like 'Armada'ya yemeğe geldik.' which means 'We came to Armada to dine.' in English shares that the person is in Armada.

366 tweets are marked and 228 of them have location privacy.

### 4.3.6 Feature 6
Feature 6 consists of 18 special verbs that use to describe where you are. For instance; 'geldim', 'geldik' and 'gitme' are the verbs that are used when you want to express where you are going or where you are. For example, 'Eve gidiyorum' which means 'I am going to the home' in English share location information that you are going to be at home very soon. Another example is; 'Annemlere geldik' which means 'We came to my mother's house' share location information.

94 tweets have feature 6 and 89 of them share location.

In Figure 3, you can see the features individually and their tweet counts for the two classes. First class stands for location sharing existence and the other stands for location sharing non-existence.

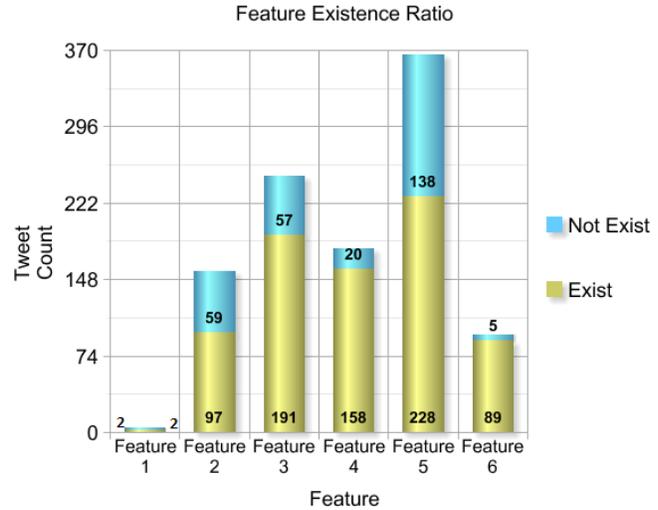

**Figure 3: Feature Existence Ratio**

## 5. Analysis on Classification Algorithms
### 5.1 Weka
Weka [15] is a free distribution software developed by Waikato University which provides tools and algorithms for data mining and predictive modelling. In this study, Weka 3.6 is used for choosing the most suitable classification algorithm among the available algorithms based on the feature classification and location privacy data.

### 5.2 Evaluating Data Mining Algorithms
#### 5.2.1 Input Data Set
Input data given to classification algorithms is a collection of rows where the attributes are feature1, feature2, feature3, feature4, feature5, feature6 and class label. The data is exported from the tweet database by SQL statements. The full dataset contains 500 tuples with 7 attributes (6 predictors and 1 class label). The class label 1 indicates the location sharing and 0 indicates the no location sharing.

#### 5.2.2 Process
73 classifiers and different test options are used in order to evaluate classifiers.

First, 73 classifiers resulting from several classification algorithms are evaluated with the test option of 10 fold cross-validation. After getting all the results, we sort them by looking at the Correctly Classified Instances percentage of the output. The top three algorithms are chosen. They are due to J48graft, J48 and END classification algorithms. In order to decide which one is the most suitable classifier, we try different test options with the same data for the three classifiers.

| Algorithm | Cross-validation | Percentage split | Correctly Classified Instances |
|---|---|---|---|
| J48graft | | 66 | 88,82 |
| J48 | | 66 | 88,23 |
| END | | 66 | 88,23 |
| J48graft | 50 | | 87,4 |
| J48graft | 10 | | 87,6 |
| J48 | | 90 | 87,2 |
| J48 | 10 | | 87,4 |
| J48 | 30 | | 87,4 |
| J48 | | 50 | 87,2 |
| J48 | 20 | | 87,2 |
| J48 | 70 | | 87,2 |
| END | 50 | | 87,2 |
| END | 10 | | 87,4 |
| J48graft | | 30 | 83,14 |

**Figure 4: Classifier comparison table**

As shown on the Figure 4 J48graft algorithm is the most accurate one when classifying our instances.

*5.2.3 Evaluation Results*

J48graft produces the decision tree bellow:

**Decision Tree:**

feature4 = 0

| feature6 = 0

| | feature3 = 0: 0 (175.0/3.0)

| | feature3 = 1

| | | feature2 = 0: 0 (55.0/17.0)

| | | feature2 = 1: 1 (43.0/15.0)

| feature6 = 1: 1 (49.0/5.0)

feature4 = 1

| feature5 = 0: 0 (7.0/1.0)

| feature5 = 1: 1 (171.0/14.0)

Since J48graft classifier gives the best result its decision tree is used in when developing the Google Chrome extension. The J48 classifier also gives the same decision tree. Different decision trees of different classifiers can be used for getting different results. For instance, the decision tree of ADTtree is shown below:

**Decision Tree:**

| (1)feature4 = 0: -0.455

| (1)feature4 = 1: 1.012

| | (3)feature5 = 0: -1.712

| | (3)feature5 = 1: 0.547

| | | (7)feature1 = 0: 0.013

| | | (7)feature1 = 1: 0.234

| | (6)feature1 = 0: -0.029

| | (6)feature1 = 1: 0.486

| (2)feature3 = 0: -0.82

| | (4)feature5 = 0: -0.977

| | (4)feature5 = 1: -0.044

| (2)feature3 = 1: 0.691

| | (5)feature5 = 0: -0.188

| | (5)feature5 = 1: 0.054

However, this decision tree is not used because of its poor results. ADTtree's correctly classified instances output is 83 percentage which is less than J48's results.

## 6. Google Chrome Extension

The Google Chrome extension developed to prevent twitter users from sharing their sensitive location information. The extension is optional and it can be activated or deactivated from its popup menu. When it is activated the extension runs the algorithm derived from the decision tree. Extension checks your tweet before clicking the 'Tweetle' button, it checks when you hover over the button. And if it predicts an implicit location sharing it will warn you with a message and an icon. As a warning the extension write 'Konum paylasiyor olabilirsiniz!' which means 'You may be sharing your location!' in English between the text area of the tweet and the header of the new tweet screen. Figure 4 is a screenshot of our program.

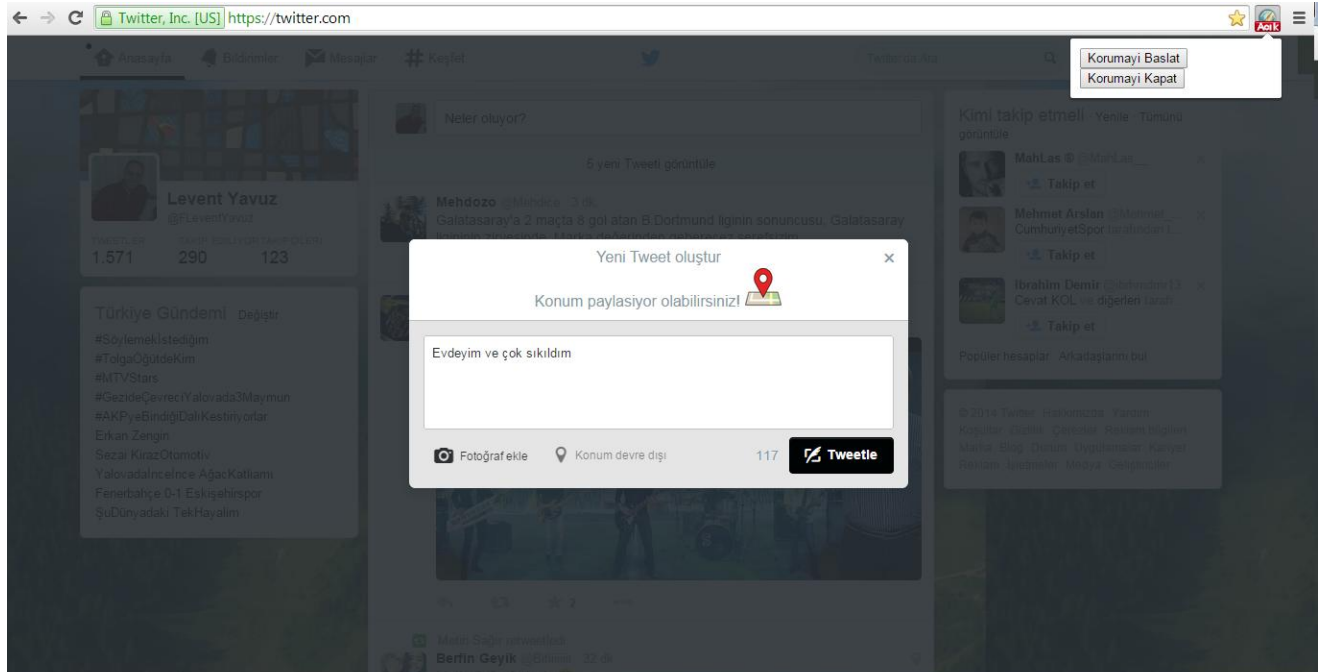

**Figure 4: Screenshot of the Google Chrome Extension**

## 7. Conclusion

The tool generated by this study could be very useful in various ways. It could help social media users to audit their location privacy sharing. It could be also useful to educate them on their location privacy. Altough the tool is implemented for Twitter, it can be easily deployed for other micro blogging social media platforms such as Facebook and MySpace.

Although the current accuracy is over 80%, future work may enhance this value by extracting new features mostly based on natural language studies. For instance, we can add more specific words to the available features but also adapt new features to help algorithm to work with more features and words based on the Turkish language understanding.

Following the same methodology one can easily extend the work to other languages such as English and French. It suffices to replace language specific features.

## 8. References

[1] Kadaba, Lini S., 2012, What is privacy? As job-seekers are judged by their tweets and Facebook posts, uncertainty abounds Retrieved form: http://articles.philly.com/2012-05-03/news/31539376_1_facebook-photos-facebook-passwords-employers

[2] Pontes, T., 2012, Beware of What You Share Inferring Home Location in Social Networks, Retrieved from: http://ieeexplore.ieee.org/xpl/articleDetails.jsp?tp=&arnumber=6406403&url=http%3A%2F%2Fieeexplore.ieee.org%2Fxpls%2Fabs_all.jsp%3Farnumber%3D6406403

[3] Foursquare, 2014, Privacy 101, Retrieved from: https://foursquare.com/privacy/

[4] Weidemann C., 2013, http://geosocialfootprint.com/

[5] Weidemann C., 2013 , Social Media Location Intelligence: The Next Privacy Battle - An ArcGIS add-in and Analysis of Geospatial Data Collected from Twitter.com, Retrieved from: http://journals.sfu.ca/ijg/index.php/journal/article/view/139

[6] Groeveneld F., Borsboom B., Amstel, B., 2011, Over-sharing and Location Awareness, Retrieved from: https://cdt.org/blog/over-sharing-and-location-awareness/

[7] Twitter, 2014, Twitter Privacy Policy, Retrieved from: https://twitter.com/privacy

[8] Taşlıoğlu, H., 2014, Irony Detection on Turkish Microblog Texts, Thesis Submitted to The Graduate School of Natural And Applied Sciences of Middle East Technical University

[9] Stefanidis, A., 2011, Harvesting ambient geospatial information from social media feeds, Retrieved from: http://www.academia.edu/1472285/Harvesting_Ambient_Geospatial_Information_from_Social_Media_Feeds

[10] http://twitter4j.org

[11] https://dev.twitter.com

[12] http://www.mysql.com/

[13] http://www.tdk.gov.tr

[14] http://www.nisanyanmap.com/?lg=t

[16] http://www.cs.waikato.ac.nz/ml/weka/